\documentstyle[twocolumn,epsf]{jpsj}

\def\runtitle{Superconducting M\"obius Strip}
\def\runauthor{Masahiko \textsc{Hayashi} and Hiromichi \textsc{Ebisawa}}

\title{%
Little-Parks Oscillation of 
Superconducting M\"obius Strip
}

\author{%
Masahiko \textsc{Hayashi}\thanks{E-mail address: 
hayashi@cmt.is.tohoku.ac.jp} and
Hiromichi \textsc{Ebisawa}
}

\inst{%
Graduate School of 
Information Sciences, Tohoku University, \\
Aramaki Aoba-ku, Sendai 980-8579, Japan
}

\recdate{June 15, 2001}

\abst{%
Little-Parks oscillation of a M\"obius strip 
(or ring equivalently) made of a superconductor 
is studied based on Ginzburg-Landau theory. 
It is shown that, if the strip is wide enough, 
a novel state appears when the number of 
magnetic flux quanta threading 
the ring is close to a half odd integer. 
As a result the shape of Little-Parks oscillation 
of critical temperature is modified. 
We estimate the free energy of this state and give the 
phase diagram of the superconducting M\"obius strip 
in a magnetic field. }

\kword{%
Superconductivity, 
Little-Parks oscillation, Topological Matter, M\"obius strip
}

\begin{document}
\sloppy
\maketitle

Recently novel type of crystals have been grown by the 
group in Hokkaido University \cite{Tanda}, 
which are referred to as \lq\lq topological matter\rq\rq. 
These crystals are ring, disk or cylinder shaped
with one of their crystal axis along the 
azimuthal direction. 
Surprisingly, even a ring of a quasi-one-dimensional 
charge-density-wave system, e.g., ${\rm NbSe}_{3}$, 
was created and the formation of the 
charge-density-wave was experimentally confirmed
\cite{Okajima}, 
which indicates that these samples are good crystals 
and relatively free from the expected disorder 
originating from the bending of the crystal axis. 

Among these crystals we concentrate on 
one of the most subtle structure, 
the M\"obius strip. 
M\"obius strip is a representative object 
which has non-trivial topology. 
Therefore it is interesting to see how its 
topological nature affects the 
physical properties of the system. 
Experimentally, the M\"obius strip of ${\rm Nb Se}_{3}$ 
has been created \cite{Tanda-moebius}. 
We especially pay attention to the case where 
${\rm Nb Se}_{3}$ behaves as a superconductor. 
In this sample, the most conducting axis 
is along the azimuthal direction. 
As we see later this is actually the best 
configuration to observe non-trivial topological 
effects in the superconducting state. 

The M\"obius geometry was previously 
investigated concerning the correlated electrons in 
mesoscopic systems. 
Persistent current through the 
so-called \lq\lq M\"obius ladder\rq\rq was 
theoretically studied by Mila, Stafford and 
Coppeni \cite{Mila}. 
The M\"obius ladder is a simplified version of 
M\"obius strip which consists of only one closed electronic 
channel winding twice along a circle. 
In case of a M\"obius strip, however, the 
situation is more complicated, 
since there are many channels along the strip. 
The aim of this letter is to clarify how the 
persistent current ($=$supercurrent) 
through the superconducting M\"obius strip 
is affected by its topological structure. 

To answer this question 
we study Little-Parks oscillation, 
which is the 
oscillation of superconducting critical 
temperature as a function of the magnetic 
flux threading the ring 
\cite{Little-Parks,Groff-Parks}. 
We study this phenomenon based on the Ginzburg-Landau (GL)
theory. 
Bearing the experimental situation in mind, 
we assume that the M\"obius ring is formed by an 
array of coupled one-dimensional chains, 
although the continuum limit is taken in the end of the 
calculation. 
Throughout this letter, the thickness of the strip is 
disregarded and the strip is assumed to be 
two-dimensional. 

\begin{figure}
\begin{center}
{\parbox{7cm}
{\epsfxsize=7cm \epsfbox{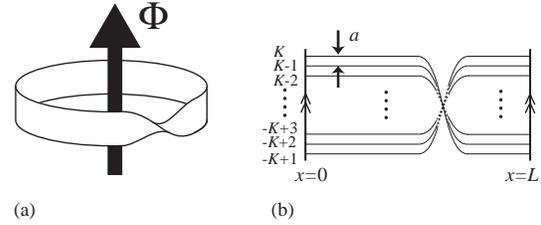}}}
\end{center}
\caption{(a) Structure of the M\"obius ring 
threaded by a magnetic flux $\Phi$. 
(b) Chains consisting the ring. 
Both ends are identified.
}
\label{moebius}
\end{figure}

Our model is described by the following GL free energy, 
\begin{eqnarray}
F&=&\sum_{i=-K+1}^{K}\int_{0}^{L}{\rm d}x\, 
\bigg[\frac{1}{2 m^{*}}\left|
\left(\frac{\hbar}{i} \partial_{x}+
\frac{e^{*}}{c}A_{x}\right)\psi_{i}
\right|^{2}
\nonumber\\
&&\phantom{aaaaaaa}
+\alpha\left|\psi_{i}\right|^{2}
+\frac{b}{2}
\left|\psi_{i}\right|^{4}\bigg]\nonumber\\
&&+\sum_{i=-K+1}^{K-1}\int_{0}^{L}{\rm d}x\, 
v \left|\psi_{i+1}-\psi_{i}\right|^{2}. 
\label{GL}
\end{eqnarray}
Here the $x$-coordinate ($0 < x \leq L$) 
is taken along the ring. 
The geometry of the system is shown in fig. \ref{moebius}. 
$c$, $m^{*}$ and $-e^{*}$ are the speed of light, and 
the mass and charge of a Cooper pair, 
respectively. 
$v$ is a parameter of interchain Josephson 
coupling, and 
$\alpha$ and $b$ are GL parameters. 
The order parameter of the $i$-th chain is 
denoted by $\psi_{i}(x)$. 
Boundary condition for the order parameter is 
given by, 
\begin{equation}
\psi_{i}(0) = \psi_{-i+1}(L), \phantom{aa}
\psi_{i}(L) = \psi_{-i+1}(0). 
\end{equation}
This comes from the fact that 
the $i$-th chain and the $(-i+1)$-th chain 
are connected, making a closed chain of 
total length $2 L$, 
which is a direct conclusion of 
the M\"obius geometry (see fig. \ref{moebius}). 
Number of chains in the cross section of the strip 
is assumed to be even ($= 2 K$) for simplicity. 
The $x$-component of vector potential $A_{x}$ 
is set to be a constant 
$A_{x}=\Phi/L$, where $\Phi$ is 
magnetic flux threading the ring. 
In this letter we consider 
only the Aharonov-Bohm flux, 
neglecting the effects of 
magnetic field applied on the strip. 

We write the order parameter as 
$\psi_{i}(x) = \chi_{i} {\rm e}^{i \theta_{i}(x)}$ and 
assume the amplitude $\chi_{i}$ to be a constant. 
The following choice of $\theta_{i}(x)$ 
may be natural, 
\begin{equation}
\theta_{i}(x) = 
\left\{
    \begin{array}{ll}
	\pi\frac{x}{L}l_{i} + \pi l_{-i+1}& (i \le 0)\\
	\pi\frac{x}{L}l_{i} & (i \ge 1)
    \end{array}
\right.,
\label{theta}
\end{equation}
where $l_{i}$'s are real numbers such that 
$(l_{i} + l_{-i+1})/2$ is an integer. 
This makes the order parameter single-valued. 
Since the $i$-th chain and the $(-i+1)$-th chain 
are essentially the same chain, we 
assume that they have same amplitude and 
wave number:  
$\chi_{i} = \chi_{-i+1}$ and $l_{i} = l_{-i+1}$. 
Then $l_{i}$'s are further limited to integers. 
By substituting these forms into 
Eq. (\ref{GL}) and performing the integration over 
$0 < x \le L$, we obtain the following expression, 
\begin{eqnarray}
    F & = & L \sum_{i=-K+1}^{K} \biggl[\frac{\hbar^{2}}{2 m^{*}}
    \left(\frac{2 \pi}{L}\right)^{2} 
    \left(\frac{l_{i}}{2}-f \right)^{2}\chi_{i}^{2}
    \nonumber \\
    && \phantom{aaaaaaaaa}
    + \alpha \chi_{i}^{2} + \frac{b}{2} \chi_{i}^{4}\biggr]
    \nonumber  \\
    && + L \sum_{i=-K+1}^{K-1}v \left(\chi_{i+1}^{2} + 
    \chi_{i}^{2} - 2\, \chi_{i+1}\,  \chi_{i} \,
    \delta_{l_{i+1} l_{i}}\right)
    \nonumber  \\
    && + 2 L v (1 - \cos \pi l_{1}) \chi_{0}\chi_{1},
    \label{free-energy}
\end{eqnarray}
where $f=\Phi/\phi_{0}$ with $\phi_{0}=hc/e^{*}$. 
Next we minimize the free energy with respect to 
$\chi_{i}$'s and $l_{i}$'s. 
In order to minimize the free energy in the bulk, 
namely the third term of the second line, 
we set $l_{i}=l$ for all $i$. 
This, however, gives rise to an increase of the 
free energy arising from the 
last term of Eq. (\ref{free-energy}), 
if $l$ is an odd number. 

Here we transfer to the continuum 
description of the free energy 
Eq. (\ref{free-energy}) assuming 
that the characteristic length scale in 
the transverse direction, namely the 
transverse coherence length, 
is much larger than interchain spacing 
which we denote by $a$. 
We define the continuous variable by $y = a i$.
After scaling the order parameter 
and the constants appropriately, 
\begin{equation}
    \frac{\chi_{i}}{\sqrt{a}} \longrightarrow \Psi(y), 
    \phantom{aaa}
    v  \longrightarrow \frac{\tilde v}{a^{2}},
    \phantom{aaa}
    b \longrightarrow \frac{b'}{a},
    \label{scaling}
\end{equation}
we obtain the continuum free energy as, 
\begin{eqnarray}
    F & = & L \int_{-W/2}^{W/2} {\rm d}y 
    \bigg[{\tilde v}(\partial_{y} \Psi)^{2} 
    + {\bar \alpha} \Psi^{2} + 
    \frac{b'}{2} \Psi^{4}\bigg]
    \nonumber  \\
    &&+ 2 L \frac{\tilde v}{a}
    (1 - \cos \pi l) \Psi(0)^{2}.
    \label{continuum}
\end{eqnarray}
The width of the strip is given by $W = 2 K a$ and 
$\bar \alpha$ is given by 
$\alpha + \frac{\hbar^{2}}{2 m^{*}}
\left(\frac{2 \pi}{L}\right)^{2}
\left(\frac{l}{2}-f\right)^{2}$. 
By setting 
$\alpha \equiv \alpha(0) (\frac{T}{T_{c}}-1)$ 
and introducing zero-temperature 
longitudinal ($\parallel$ chains) 
and transverse ($\bot$ chains) coherence length by 
$\xi_{\parallel}(0) \equiv \hbar^{2}/(2 m^{*} \alpha(0))$ 
and 
$\xi_{\bot}(0) \equiv {\tilde v}/\alpha(0)$, 
respectively, 
Eq. (\ref{continuum}) can be simplified to, 
\begin{eqnarray}
    F & = & L \alpha(0) \int_{-W/2}^{W/2}{\rm d}y\, 
    \bigg[\xi_{\bot}(0)^{2}(\partial_{y} \Psi)^{2}
    +({\bar t}-1) \Psi^{2} + \frac{\tilde b}{2}\Psi^{4}
    \nonumber  \\
    & & + \frac{\xi_{\bot}(0)^{2}}{a} (1 - \cos \pi l)
    \delta(y) \Psi^{2}\bigg],
    \label{continuum2}
\end{eqnarray}
where ${\bar t}$ is given by 
$t + (2 \pi \xi_{\parallel}(0)/L)^{2}(l/2 - f)^{2}$ 
with $t \equiv T/T_{c}$ and 
${\tilde b} = b'/\alpha(0)$. 
Here $T_{c}$ denotes the mean field critical temperature 
without magnetic flux. 
In the last line of Eq. (\ref{continuum}) and 
Eq. (\ref{continuum2}), 
$a$ is kept to handle the diverging constant 
in the limit of $a \rightarrow 0$. 

The solution for $\Psi(y)$ is obtained by solving 
the equation, 
\begin{eqnarray}
    \frac{\delta F}{\delta \Psi} & = & 
    2 L \alpha(0) \bigg[- \xi_{\bot}(0)^{2} \Psi''
    + ({\bar t}-1) \Psi + {\tilde b} \Psi^{3}
    \nonumber  \\
    && + \frac{2 \xi_{\bot}(0)^{2}}{a} ( 1 - \cos \pi l)
    \delta(y) \Psi\bigg]
    \nonumber\\
    &=& 0. 
    \label{GL-eqn}
\end{eqnarray} 
Because of the assumption made just below 
Eq. (\ref{theta}), 
only the symmetric solution $\Psi(y) = \Psi(-y)$ is 
allowed. 
We should note here that the term including 
delta function is nonzero only when 
$l$ is an odd number. 
Therefore the characteristics of the 
solutions for even-$l$ and odd-$l$ 
differ. 

It is easy to see that the even-$l$ states are 
ordinary \lq\lq Little-Parks states\rq\rq, where 
the phase of the order parameter of each chain 
changes by a multiple of 
$2 \pi$ when one goes around the ring once. 
The critical temperature of each even-$l$ state 
is determined from ${\bar t}-1=0$ and the free energy 
of the ordered state is obtained as  
$F = - L W \alpha(0) ({\bar t}-1)^{2}/(2 {\tilde b})$ 
after a simple calculation. 

The odd-$l$ states are essentially different. 
If we go around the ring once 
the phase increases by an odd number times 
$\pi$ and, 
to compensate the frustration caused by 
this \lq\lq fractional winding number\rq\rq, 
a phase shift of $\pi$ occurs 
on the line $y=0$, 
which is actually the origin of the 
$\delta$-function in Eq. (\ref{GL-eqn}). 
The solution for odd-$l$ state is obtained by 
solving the ordinary 
GL equation (without the $\delta$-function term
in Eq. (\ref{GL-eqn})) in the region 
$y > 0$ and $y < 0$ separately, 
and connecting them 
at $y = 0$. 
The integration of 
Eq. (\ref{GL-eqn}) from $x = -\epsilon$ 
to $x = \epsilon$ with 
$\epsilon$ being a positive 
infinitesimal number
yields,  
\begin{equation}
    \lim_{\epsilon\longrightarrow 0}
    \left\{\Psi'(\epsilon) - \Psi'(-\epsilon) 
    \right\} = 4 \frac{\Psi(0)}{a}, 
\label{connect}
\end{equation}
which gives the condition of the connection
\cite{Greiner}. 

\begin{figure}
\begin{center}
{\parbox{7cm}
{\epsfxsize=7cm \epsfbox{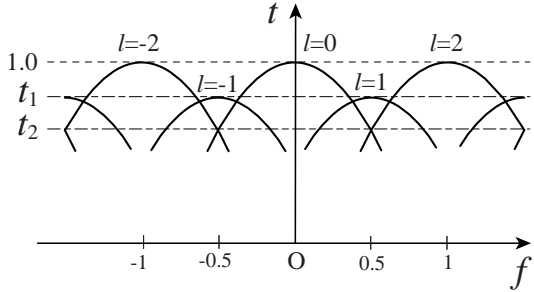}}}
\end{center}
\caption{
Critical temperature of each 
state labelled by $l$. 
The maximum of the critical temperature 
of the odd-$l$ state is indicated by $t_{1}$, 
whereas the minimum of that of 
even-$l$ state 
is indicated by $t_{2}$. 
}
\label{phase-diag-1}
\end{figure}

Next we estimate the critical temperature of 
the odd-$l$ state. 
Since near the critical temperature
the amplitude of the order parameter is small 
and the coherence length is large, 
we may introduce a variational solution  
$\Psi = \eta (|y| + \frac{a}{2})$ 
satisfying Eq. (\ref{connect}) with 
$\eta$ being a variational parameter. 
By substituting this expression into 
Eq. (\ref{continuum2}) and performing the 
integration over $-W/2 \le y \le W/2$, 
we obtain the variational free energy. 
The last term of Eq. (\ref{continuum2}) 
vanishes in the limit of $a \rightarrow 0$. 
The critical temperature 
is estimated as a temperature below which 
the solution with nonzero $\eta$ exists. 
Namely, 
${\bar t} - 1 + 12 (\xi_{\bot}(0)/W)^{2} = 0$
determines the critical temperature. 
By comparing this with 
the critical temperature 
of even-$l$ state, we can determine 
the phase diagram near $T_{c}$, 
which is schematically shown in 
fig. \ref{phase-diag-1}. 
As one can see from this figure, 
the odd-$l$ state is realized only 
when $t_{1}>t_{2}$ is satisfied, 
where $t_{1}$ and $t_{2}$ are the maximum 
of the critical temperature 
of odd-$l$ state and
the minimum of that of 
even-$l$ state, respectively, 
as shown in fig. \ref{phase-diag-1}. 
Here we introduce new parameters, 
$r_{\bot} = \xi_{\bot}(0)/W$ 
and $r_{\parallel} = \xi_{\parallel}(0)/L$. 
Since $t_{1} = 1 - 12 r_{\bot}^{2}$ and 
$t_{2} = 1 - \pi^{2} r_{\parallel}^{2}$ 
according to the present calculation, 
we obtain the criterion for the odd-$l$ state 
as $r_{\bot}<\pi r_{\parallel}/(2 \sqrt{3})$. 

Next we study the low temperature region 
$t \ll 1$ of odd-$l$ state. 
As a solution of Eq. (\ref{GL-eqn}), 
we adopt the following formula, 
which satisfies $\lim_{y\rightarrow \pm\infty}
\Psi(y)={\rm const.}$, 
\begin{eqnarray}
    \Psi(y) & = & \sqrt{\frac{1-{\bar t}}{\tilde b}}
    \tanh \left(\sqrt{\frac{1-{\bar t}}{2}}
    \frac{|y|+d}{\xi_{\bot}(0)}
    \right),
    \label{solution}
\end{eqnarray}
where $d$ is determined so that 
Eq.(\ref{connect}) is satisfied. 
Here we have neglected the effect of 
the boundaries $y = \pm W/2$ 
assuming $\xi_{\bot}(0) \ll W$. 
The condition for $d$ reduces to, 
\begin{equation}
    \sqrt{\frac{1-{\bar t}}{2}}
    \frac{a}{\xi_{\bot}(0)}=
    \sinh 
    \left(\sqrt{\frac{1-{\bar t}}{2}}
    \frac{2 d}{\xi_{\bot}(0)}
    \right),
    \label{connection}
\end{equation}
which is fulfilled by setting $d \approx a/2$ 
because $a \ll \xi_{\bot}(0)$ is supposed. 
The free energy is calculated as follows. 
By substituting Eq. (\ref{solution}) 
into Eq. (\ref{continuum2}) we obtain the following 
expression, 
\begin{eqnarray}
    &&F \simeq - L W 
    \frac{\alpha(0)(1 - {\bar t})^{2}}{2{\tilde b}}
    \left[1 - \frac{8 \sqrt{2}}{3}
    \frac{r_{\bot}}{\sqrt{1-{\bar t}}}
    \right],\phantom{aa}(r_{\bot} \ll 1). 
    \label{odd-l-energy}
\end{eqnarray}
Again we have taken 
$a \rightarrow 0$ limit here and the last term of 
Eq. (\ref{continuum2}) vanished. 

Next we determine the phase boundary 
between even-$l$ and odd-$l$ state. 
The free energy of even-$l$ and odd-$l$ state, 
normalized by 
$L W \alpha(0)/(2 {\tilde b})$, 
are given respectively by 
\begin{eqnarray}
    f_{l}^{({\rm e})}
    &=& -(1-t)^{2} + 2 (1-t) (2 \pi r_{\parallel})^{2}
    \left(\frac{l}{2}-f\right)^{2} + 
    {\cal O}(r_{\parallel}^{4}),
    \label{even}  \nonumber\\
    &&\\
    f_{l}^{({\rm o})}
    &=& -(1-t)^{2} + 2 (1-t) (2 \pi r_{\parallel})^{2}
    \left(\frac{l}{2}-f\right)^{2}
    \nonumber \\
    && +\frac{8 \sqrt{2}}{3}(1 - t)^{3/2} r_{\bot} 
    + {\cal O}(r_{\parallel}^{4},r_{\bot}^{2},
    r_{\parallel}^{2}r_{\bot}). 
    \label{odd}
\end{eqnarray}
By equating $f_{2k}^{({\rm e})}$ and $f_{2k + 1}^{({\rm o})}$ 
($k$ is an integer) 
we obtain the boundary between 
even-$l$ and odd-$l$ state. 
For example, the boundary between $l=0$ and 
$l=1$ state is given by 
$\frac{1}{4} + \frac{\sqrt{2}}{3\pi^{2}}
\frac{r_{\bot}}{r_{\parallel}^{2}}
\times\sqrt{1-t} = f$. 
From this result, we can  
see that, at $T=0$, the odd-$l$ state can exist only when 
$r_{\bot} < 3\pi^{2}r_{\parallel}^{2}/(4 \sqrt{2})$
is satisfied, 
since otherwise the boundary between $l=0$ and $l=1$
state crosses 
the boundary between $l=1$ and $l=2$ 
state at nonzero temperature, 
giving rise to direct transition 
between $l=0$ and $l=2$ state 
below the crossing temperature. 

\begin{figure}
\begin{center}
{\parbox{7cm}
{\epsfxsize=7cm \epsfbox{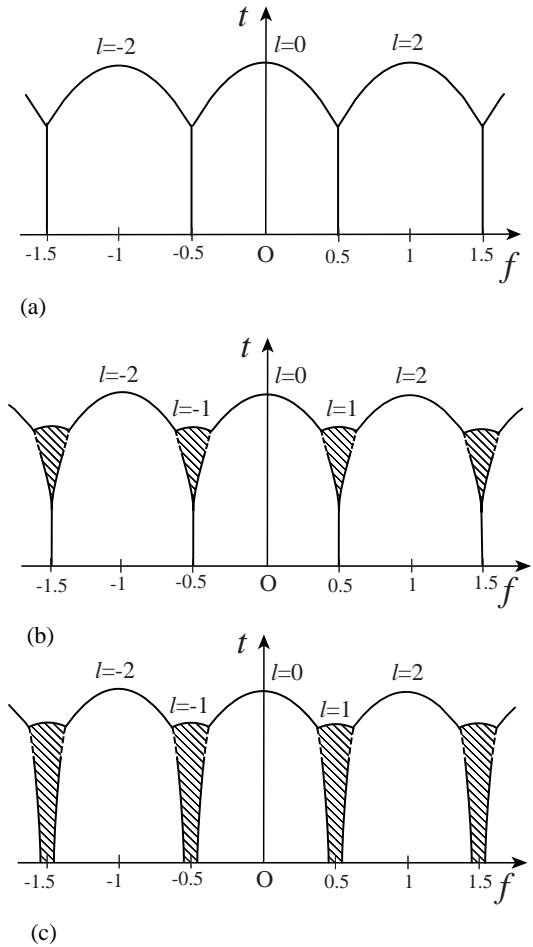}}}
\end{center}
\caption{
Phase diagram of the M\"obius ring 
for the case of (a) $\frac{\pi}{2 \sqrt{3}} 
r_{\parallel} <
r_{\bot}$, 
(b) 
$\frac{3\pi^{2}}{4 \sqrt{2}}
r_{\parallel}^{2} <
r_{\bot} <
\frac{\pi}{2 \sqrt{3}} 
r_{\parallel}$ 
and 
(c) $r_{\bot} <
\frac{3\pi^{2}}{4 \sqrt{2}}
r_{\parallel}^{2}$. 
Odd-$l$ states appears in the hatched 
region. 
}
\label{phase-diag-2}
\end{figure}

The obtained phase diagram is 
given in fig. \ref{phase-diag-2}. 
Solid lines show the phase boundaries 
obtained from the above argument. 
The broken lines show guess of the 
phase boundaries in the intermediate 
temperature region, 
where the present treatment  
is not applicable. 
As we see from the figure, 
there can be three types of behaviors 
depending on the geometry of the system. 
As is pointed out above, the odd-$l$ state 
are not at all stable when 
$\frac{\pi}{2 \sqrt{3}} 
r_{\parallel} < 
r_{\bot}$. 
In this case we obtain the situation 
depicted in fig. \ref{phase-diag-2} (a). 
If $\frac{3\pi^{2}}{4 \sqrt{2}}
r_{\parallel}^{2} <
r_{\bot} <
\frac{\pi}{2 \sqrt{3}} 
r_{\parallel}$ 
is satisfied, the odd-$l$ state appears 
though it is not yet stable at  
zero temperature 
and we obtain a phase diagram like 
fig. \ref{phase-diag-2} (b). 
If $r_{\bot} <
\frac{3\pi^{2}}{4 \sqrt{2}}
r_{\parallel}^{2}$ 
is satisfied, the odd-$l$ state 
survives until zero temperature is reached, 
as depicted in fig. \ref{phase-diag-2} (c). 
In this letter
we have assumed that 
$r_{\parallel}$ is small so that 
$\frac{3\pi^{2}}{4 \sqrt{2}}
r_{\parallel}^{2} <
\frac{\pi}{2 \sqrt{3}} 
r_{\parallel}$ is 
satisfied. 
In the actual M\"obius crystal 
fabricated by Tanda and coworkers \cite{Tanda-moebius}, 
$\xi_{\parallel}(0)$ is much larger than 
$\xi_{\bot}(0)$ because of the sample anisotropy, 
and basically the situation (b) or (c) is 
rather easily realized even if the strip width is 
not so large. 

\begin{figure}
\begin{center}
{\parbox{7cm}
{\epsfxsize=7cm \epsfbox{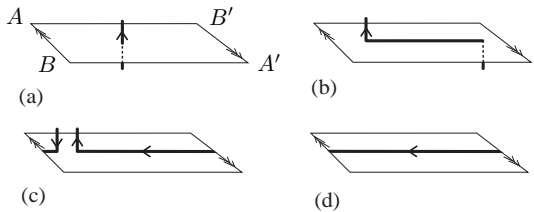}}}
\end{center}
\caption{
Formation process of the odd-$l$ state 
with a $\pi$-phase shift line in the 
midst of the M\"obius strip. 
The strip is shown in the expanded form. 
The line ${\rm AB}$ should be identified 
with ${\rm A'B'}$. 
The bold solid line indicates a vortex line. 
The vortex line is drawn also in the outside of the 
strip for the sake of clarity. 
(a) A vortex is penetrating the strip. 
(b) The vortex is stretched in the strip to 
form an in-plane vortex segment. 
(c) The segment is almost encircling the ring. 
(d) By annihilating the out-of-plane 
vortex segments, the in-plane segment forms a 
closed loop encircling the M\"obius ring. 
}
\label{vortex}
\end{figure}

As is clarified in the above argument, 
the odd-$l$ state includes a 
$\pi$-phase shift line 
in the midst of the strip. 
This line can be viewed as a  
vortex line confined in the strip, 
which we call in-plane vortex line. 
In fig. \ref{vortex} (a) - (d), we have schematically 
depicted how an in-plane vortex line is formed. 
It is interesting to see that the in-plane 
vortex line is formed from the ordinary 
vortex penetrating the strip (fig. \ref{vortex} (a)), 
which appears 
in the intermediate state of the transition between 
two even-$l$ states, a phenomenon usually called the 
\lq\lq phase slip\rq\rq. 
Then the state (a) must have larger free energy than the 
two even-$l$ states before and after the phase slip.  
From this we can conclude that when the odd-$l$ state (d) 
is more stable than even-$l$ states, it must 
be more stable than the state (a). 
Although it is not clear whether or not the odd-$l$ state 
can be, for a certain $f$, 
the most stable configuration 
among all the possible configurations 
of the order parameter, 
we consider that the odd-$l$ state 
is a strong candidate because of its uniformity 
along the strip. 
Further investigation of this point, 
however, is left for the future studies. 

Our results are experimentally accessible 
using topological matter \cite{Tanda-moebius}. 
However, one should note that 
the Little-Parks oscillation is actually observable only 
for the samples with a diameter less than  
1 $\mu$m \cite{Tinkham}. 
In such systems, effect of thermal fluctuation is 
not negligible \cite{Kanda}. 
This effect is most significant when the number of 
the magnetic flux quanta in the ring is close to 
a half odd integer \cite{Hayashi-Ebisawa} 
and, therefore, it is serious for 
the odd-$l$ states. 
Because of this, the condition for the observation of 
odd-$l$ state can be severer in actual systems. 
To overcome this point, 
further development of fabrication 
and manipulation technique of the topological matter 
is required. 

In summary, we have investigated the 
superconducting M\"obius strip under 
a magnetic field. 
We have pointed out a possibility of 
a new state which appears when 
the number of magnetic 
flux quanta in the ring is close to a half odd integer 
and clarified its structure. 
The magnetic phase diagram is 
obtained for systems with various 
radius and strip width. 

The authors are grateful to Prof. S. Tanda and Dr. A. Kanda 
for useful discussions.

\vfill\eject

\vfill

\end{document}